# Advanced Hybrid Deep Learning Model for Enhanced Classification of Osteosarcoma Histopathology Images

Arezoo Borji[1,2,3], Gernot Kronreif[1], Bernhard Angermayr[2,4], Sepideh Hatamikia[2,1]

[1] *Austrian Center for Medical Innovation and Technology, Wiener Neustadt, Austria*

[2] *Department of Medicine, Danube Private University, Krems an der Donau, Austria*

[3] *Department of Medical Physics and Biomedical Engineering, Medical University of Vienna, Vienna, Austria*

[4] *Patho im zentrum, St. Pölten, Austria*

**Abstract**

Recent advances in machine learning are transforming medical image analysis, particularly in cancer detection and classification. Techniques such as deep learning, especially convolutional neural networks (CNNs) and vision transformers (ViTs), are now enabling the precise analysis of complex histopathological images, automating detection, and enhancing classification accuracy across various cancer types. This study focuses on osteosarcoma (OS), the most common bone cancer in children and adolescents, which affects the long bones of the arms and legs. Early and accurate detection of OS is essential for improving patient outcomes and reducing mortality. However, the increasing prevalence of cancer and the demand for personalized treatments create challenges in achieving precise diagnoses and customized therapies.

We propose a novel hybrid model that combines convolutional neural networks (CNN) and vision transformers (ViT) to improve diagnostic accuracy for OS using hematoxylin and eosin (H&E) stained histopathological images. The CNN model extracts local features, while the ViT captures global patterns from histopathological images. These features are combined and classified using a Multi-Layer Perceptron (MLP) into four categories: non-tumor (NT), non-viable tumor (NVT), viable tumor (VT), and none-viable ratio (NVR). Using the Cancer Imaging Archive (TCIA) dataset, the model achieved an accuracy of 99.08%, precision of 99.10%, recall of 99.28%, and an F1-score of 99.23%. This is the first successful four-class classification using this dataset, setting a new benchmark in OS research and offering promising potential for future diagnostic advancements.

**Keywords:** Osteosarcoma; vision transformer; histopathology; feature extraction.

## 1. Introduction

Osteosarcoma is recognized as an aggressive form of bone cancer that commonly affects adolescents and children [1]. To determine the optimal treatment and assess the percentage of tumor necrosis, it is crucial to examine various histological regions [2]. However, traditional diagnostic methods, which rely heavily on manual examination of histopathological slides, are time-consuming, prone to observer bias, and often limited in diagnostic precision. Given the increasing prevalence of cancer and the demand for personalized treatments, there is an urgent need for automated, efficient, and accurate diagnostic tools [3]. Pathology

Corresponding author: Dr. Sepideh Hatamikia; email address: Sepideh.Hatamikia@dp-uni.ac.at

informatics, a rapidly expanding field within medical informatics, aims to extract valuable insights from medical pathology data. In recent years, digital pathology has experienced significant growth, with histopathological image analysis playing a vital role in the diagnosis and classification of OS [4]. Machine learning (ML) techniques, particularly deep learning (DL), have become increasingly prominent in histology image classification and segmentation [5]. ML methods, including neural networks, are proving to be highly effective in classifying and analyzing images of various cancers [6]. Several studies have focused on extracting a broad set of features, not all of which are necessarily relevant. For instance, Yu et al. [7] extracted over 9,000 features from images, covering aspects such as color, texture, object identification, granularity, and density. Irshad et al. [8] explored various image analysis techniques, including thresholding based on region growth, k-means clustering, and morphological features such as area and shape structures. Arunachalam et al. [9] introduced a method that utilized multi-level thresholding and shape segmentation to identify viable tumors, necrotic regions, and non-tumor areas in OS histology slides. Similarly, Malon et al. [10] trained a neural network to classify mitotic and non-mitotic cells based on morphological features like color, texture, and shape. However, many of these methods primarily emphasize nuclei segmentation rather than direct classification of tumor or non-tumor regions.

The advent of DL, particularly convolutional neural networks (CNNs), has significantly advanced computer vision and pattern recognition in histopathology [11]. CNNs have shown great promise in extracting key local features from images, including edges and textures. Studies by Litjens [12] and Spanhol et al. [13] demonstrated the effectiveness of CNNs in breast image classification. CNNs typically extract features through convolutional layers and classify these features through fully connected layers. For instance, Su et al. [14] used a fast-scanning CNN for breast cancer classification, while Spanhol et al. [13] extended the existing AlexNet architecture for various breast cancer segmentation tasks. Despite the success of CNNs, their reliance on local features limits their ability to capture global patterns in complex images such as histopathology slides of OS.

DL techniques have also been applied to OS classification using histological images in a few studies. Asmaria et al. [15] developed a CNN model to classify cell viability in H&E-stained OS datasets by employed data augmentation techniques to improve model performance. Sharma et al. [16] investigated various edge detection methods and evaluated the effectiveness of different feature sets, including Histogram of Oriented Gradients (HOG), using random forest and support vector machine (SVM) classifiers. Barzekar et al. [17] developed a new CNN structure (C-Net) specifically designed for classifying OS histological images. Hardie et al. [18] applied CNN models to detect OS, achieving an accuracy of 90.36%. This research suggested exploring more advanced DL architectures, such as Xception, to enhance diagnostic accuracy.

However, the studies also showed some limitations. For example, in Asmaria et al. [15], the classification of osteosarcoma histological images involved a segmentation step, where regions of interest were isolated before the deep learning model could be applied. This segmentation step adds complexity, increasing computational time and slowing down the overall process. Additionally, it risks losing critical global context by focusing only on specific regions, potentially leading to incomplete classifications, particularly when tumor heterogeneity plays a role. Moreover, segmentation often requires manual intervention, which introduces the possibility of human error and bias, especially when tumor boundaries are ambiguous.

Similarly, Sharma et al. [16] relied heavily on edge detection and segmentation techniques that may overlook subtler image features necessary for accurate classification. While effective for extracting well-defined structures, these methods may miss less obvious characteristics within the tumor tissue, reducing the model's ability to fully capture the complexity of the histological images.

Barzekar et al. [17] and Hardie et al. [18] also faced challenges with segmentation-based approaches. Although they achieved reasonable accuracy, their reliance on manually segmented data could lead to inconsistent results and increased variability due to human interpretation. Furthermore, segmentation can miss critical global features that contribute to tumor classification, especially in heterogeneous tumors, where subtle patterns or transitions across tissue regions are essential for an accurate diagnosis. Recent studies have also compared various DL models and hybrid approaches for cancer diagnosis. Vezakis et al. [19] compared various deep learning models for osteosarcoma diagnosis from histopathological images, finding that smaller models like MobileNetV2 outperform larger ones due to better generalization on limited data. This finds the importance of model selection to improve diagnostic accuracy and efficiency in medical imaging. Astaraki et al. [20] compared radiomics and DL approaches for predicting malignancy in lung nodules, concluding that hybrid models combining traditional radiomics and DL methods yielded the best diagnostic results. Additionally, Wang et al. [21] developed an hybrid AI-based tool, OS Histological Imaging Classifier (OSHIC), which uses digital pathology to predict OS recurrence and survival based on nuclear morphological features. To further improve medical image segmentation while preserving spatial information, Erickson et al. [22] introduced a novel DL architecture called INet, which CNNs with attention mechanisms, forming a hybrid model that integrates feature extraction and selective focus on important image regions. This hybrid approach enhanced the model's ability to accurately classify medical images by blending the strengths of both CNNs and attention-based networks, making it particularly effective for complex medical datasets. The hybrid models have the advantage of leveraging CNNs' ability to capture fine-grained local features and attention mechanisms' capacity to model long-range dependencies and global patterns within images. This allows for a more comprehensive understanding of the image, making hybrid models especially suited for tasks requiring both detailed and broad image analysis, which is why we applied this approach in our work.

This paper introduces a new hybrid DL method for classifying tumor types in OS (non-tumor (NT), non-viable tumor (NVT), viable tumor (VT), and none-viable ratio (NVR)) using H&E-stained histopathological images of OS sourced from the Cancer Imaging Archive (TCIA). The proposed hybrid approach combines CNN, Vision Transformer (ViT), and multi-layer perception (MLP) models applied directly to histological images without a segmentation step, capturing both local and global image features.

## 2. Methodology

### General experimental design

In this study, different DL techniques including ResNet, ViT, CNN, and a hybrid CNN-ViT architecture are used for classifying OS histopathological images. These models utilize both local and global feature extraction techniques to accurately classify OS tissues, and the hybrid model combines the strengths of CNNs for capturing local patterns and ViTs for modeling global context. Through this approach, we aim to establish an efficient and accurate system for histopathological image classification in OS that enhances diagnostic capabilities in cancer research. We have divided the dataset into training, validation, and test sets for all the models. We used the training set, comprising 60% of the data, to update the model's weights, and used the validation set, accounting for 15% of the data, to provide feedback on the model's generalization during training. We reserved the remaining 25% of the data for testing the final model. We have conducted all the research methods in a Google Colab environment using a Tesla T4 GPU and an Intel (R) Core (TM) i7-4790K CPU running at 4.00 GHz with 16 GB of RAM.

## 2.1. The dataset description

In this study, we have used an open-source osteosarcoma histology image dataset from the Cancer Imaging Archive (TCIA) https://www.cancerimagingarchive.net/collection/osteosarcoma-tumor-assessment/, compiled by clinical scientists at the University of Texas Southwestern Medical Center at Children's Medical Center in Dallas from 1995 to 2015. For research purposes, the dataset, which is publicly available on the TCIA website, consists of 1144 histopathological images in JPG format. We categorize the images into four classes: 1. non-tumor (NT), 2. non-viable tumor (NVT), 3. viable tumor (VT), and non-viable ratio (NVR). The NT category is the largest, with 536 images showing normal bone tissue, blood vessels, and cartilage. The categories of NVT, VT, and NVR are smaller, with 263, 292, and 53 images, respectively. Figure 1 presents sample images from each of the four categories.

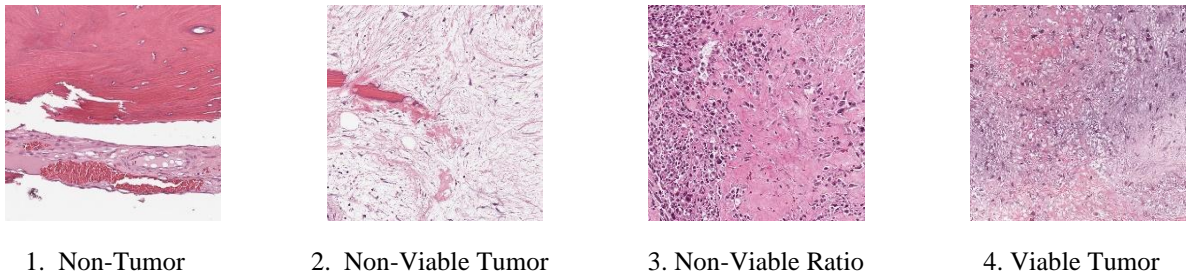

   1. Non-Tumor           2. Non-Viable Tumor        3. Non-Viable Ratio          4. Viable Tumor

Figure 1. Four samples from the dataset: (1) NT, (2) NVT, (3) NVR, and (4) VT

The definitions of all these categories are explained below:

**Non-Tumor:** Non-Tumor (NT) refers to tissue that remains unaffected by cancer. This could be surrounding healthy tissues such as muscles, bones, or organs near the tumor site, but these are not part of the cancerous mass.

**Non-Viable Tumor:** The term non-viable tumor (NVT) tissue refers to dead or necrotic tumor cells. This tissue is no longer living or functional, often due to previous treatment, such as chemotherapy, radiation, or spontaneous cell death. Non-viable tumor tissue does not have the capability to grow or spread.

**Viable Tumor:** Viable tumor (VT) tissue consists of living, active cancer cells that are capable of growth and division. This tissue poses a significant risk due to its potential to proliferate and metastasize to other parts of the body.

**Non-Viable Ratio:** The Non-viable ratio (NVR) describes the proportion of living tumor cells to dead tumor cells in a specific area. A lower viable-to-non-viable ratio indicates that treatment has successfully killed more of the tumor, whereas a higher ratio indicates that a significant portion of the tumor remains active and potentially dangerous.

## 2.2. Image pre-processing

The TCGA images (Section 2.1), were resized to 128×128 for model processing. The dataset is imbalanced, with the NT class having the most samples (536 images) and the NVR class having the fewest (53 images). To address this imbalance, we have applied class weighting during training to ensure that the model pays adequate attention to the minority classes. Moreover, we have normalized them using the mean and standard deviation values.

### 2.3. ResNet

ResNet50 is a 50-layer DL model that uses residual connections to train deep networks [23]. These residual connections solve the problem of vanishing gradients, which commonly affect deep networks by allowing information to skip layers during the forward and backward pass. This allows deeper networks like ResNet50 to learn effectively, even with many layers, without degrading the model's performance. ResNet50 divides its architecture into several stages, each containing a series of convolutional layers, batch normalization, activation functions (typically ReLU), and residual connections. The network extracts hierarchical features from images, starting with low-level features like edges and textures and progressing to more complex features representing the structure and patterns within the images. By leveraging the pre-trained ImageNet weights and fine-tuning the final fully connected layer, we can adapt this deep network to our specific classification task. The combination of careful hyperparameter tuning, data augmentation, and monitoring through Tensor Board resulted in a model that achieved high accuracy and generalization. This approach underscores the potency of transfer learning in medical image analysis, enabling the adaptation of pre-trained deep networks to address domain-specific issues with minimal alteration and outstanding outcomes.

In Table 1, we have shown the architecture of ResNet50, including the layer type, input/output shapes, operations, and purpose at each stage of the network. This table specifically corresponds to the ResNet50 architecture, which is adapted for four-class classification in our study.

Table 1. The architecture of ResNet50 for four classification of OS patients

| Layer Name | Layer Type | Input Shape | Output Shape | Description |
| --- | --- | --- | --- | --- |
| Input | Image | (128, 128, 3) | (128, 128, 3) | The input image has 128×128 pixels and 3 channels (RGB). |
| Conv1 | Convolution (7×7, stride 2) | (128, 128, 3) | (64, 64, 64) | The first convolutional layer applies 64 filters with a 7×7 kernel size and stride of 2. |
| MaxPool1 | Max Pooling (3×3, stride 2) | (64, 64, 64) | (32, 32, 64) | Max pooling layer that down samples the input by taking the maximum value over 3×3 patches. |
| Residual Block 1 | Residual Block (3 layers) | (32, 32, 64) | (32, 32, 256) | First set of residual blocks. Includes 1×1, 3×3, and 1×1 convolutions, with 64 filters. |
| Residual Block 2 | Residual Block (4 layers) | (32, 32, 256) | (16, 16, 512) | The second set of residual blocks includes 1x1, 3×3, and 1x1 convolutions with 128 filters. |
| Residual Block 3 | Residual Block (6 layers) | (16, 16, 512) | (8, 8, 1024) | The third set of residual blocks includes 1×1, 3×3, and 1×1 convolutions with 256 filters. |
| Residual Block 4 | Residual Block (3 layers) | (8, 8, 1024) | (4, 4, 2048) | The fourth set of residual blocks includes 1×1, 3×3, and 1×1 convolutions with 512 filters. |
| Global Average Pooling (GAP) | Global Average Pooling | (4, 4, 2048) | (2048) | Reduces the spatial dimensions (4×4) to a 1D vector of 2048 features by averaging across each filter. |
| Fully Connected Layer | Fully Connected (Linear) | (2048) | (4) | A fully connected layer customized to output 4 class probabilities (for 4 classes). |
| SoftMax Layer | SoftMax Activation | (4) | (4) | Converts the raw logits into probabilities for each of the 4 classes (NT, NVT, VT, NVR). |

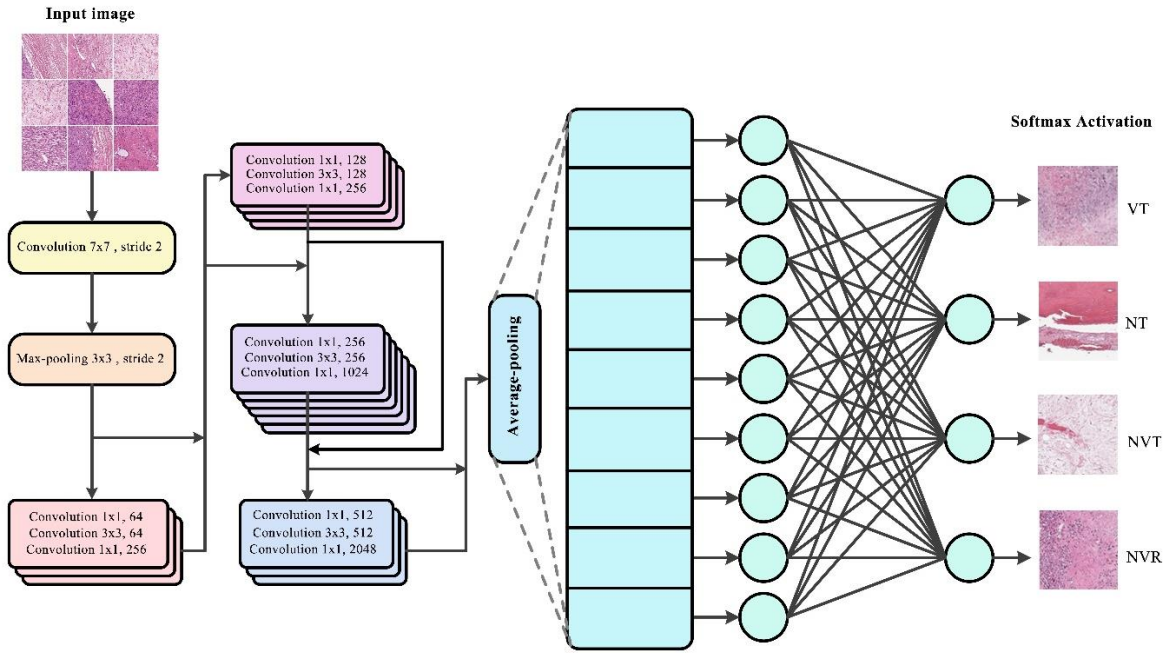

Figure 2. The structure of ResNet 50 model is used in this study.

Figure 2 illustrates the architecture of the ResNet-50 model used in this study, which consists of 50 layers including residual blocks that allow for efficient gradient flow. The model incorporates skip connections to prevent vanishing gradients, enhancing performance on complex image classification tasks by learning deeper features without degradation.

### 2.4. ViT model

The use of a ViT model for histopathological image classification represents a novel and effective approach in medical imaging analysis [24]. In this context, the ViT model processes input images by dividing them into non-overlapping patches, embedding these patches into vectors, and feeding the embedding into a series of transformer blocks. The architecture of the ViT model is based on self-attention mechanisms [25] that allows the model to focus on different parts of the image simultaneously, capturing dependencies between distant regions of the image. This is particularly important in medical imaging tasks, where subtle features distributed across the image may be critical to accurate classification. The ViT model can learn both local and global relationships by using multi-head self-attention layers to process the whole image context [26]. To adapt the ViT model for this specific classification task, the pre-trained classification head, originally designed for 1,000 ImageNet classes, was replaced with a custom head tailored for the four-class output, i.e. including NT, NVT, VT, and NVR. The modified classification layer uses a fully connected layer that outputs four probabilities, corresponding to the likelihood that the input image belongs to each class. We have connected the modified head layer to the ViT's transformer blocks for this task, enabling the model to generate predictions tailored to the histopathological data.

We have used a batch size of 32 images during training to balance computational efficiency with the ability to capture meaningful gradients during optimization. The training was conducted over twenty epochs, based on observed convergence behavior and available computational resources. We used cross-entropy loss as the loss function, measuring the difference between predicted class probabilities and true labels. The Adam optimizer was employed with parameters $\beta 1 = 0.9$ and $\beta 2 = 0.999$, which helped to stabilize the training. The optimizer, using backpropagation, adjusted the model's weights to minimize this loss. We evaluated the model's validation loss and accuracy at the end of each epoch, saving the model with the lowest validation loss as the best-performing model. This early stopping mechanism ensured that the model would not overfit the training data and would generalize well to unseen data. After completing the model training, we evaluated the final model on the test dataset to gauge its performance on unseen images. We computed standard metrics such as accuracy, precision, recall, and F1-score for each class. These metrics provided a comprehensive understanding of the model's classification performance, not only in terms of overall accuracy but also in its ability to correctly identify each class.

In Table 2, we have shown the architecture of the ViT model used in this paper for four classification of OS patients.

Table 2. The architecture of ResNet50 for four classification of OS patients

| Layer (Type) | Output Shape | The number of parameters | Layer (Type) | Output Shape | The number of parameters |
|---|---|---|---|---|---|
| Conv2d | [-1, 768, 14, 14] | 590592 | LayerNorm | [-1, 197, 768] | 1536 |
| Identity | [-1, 196, 768] | 0 | Linear | [-1, 197, 3072] | 2360064 |
| PatchEmbed | [-1, 196, 768] | 0 | Dropout | [-1, 197, 3072] | 0 |
| Dropout | [-1, 197, 768] | 0 | Linear | [-1, 197, 3072] | 0 |
| Identity | [-1, 197, 768] | 0 | Dropout | [-1, 197, 768] | 0 |
| LayerNorm | [-1, 197, 768] | 1536 | MLP | [-1, 197, 768] | 0 |
| Linear | [-1, 197, 2304] | 1771776 | Block | [-1, 197, 768] | 0 |
| Identity | [-1, 12, 197, 64] | 0 | Layer Norm | [-1, 197, 768] | 1536 |
| Linear | [-1, 197, 768] | 590592 | Linear | [-1, 197, 2304] | 1771776 |
| Dropout | [-1, 197, 768] | 0 | Identity | [-1, 12, 197, 64] | 0 |
| Attention | [-1, 197, 768] | 0 | Linear | [-1, 197, 768] | 590592 |
| LayerNorm | [-1, 197, 768] | 1536 | Dropout | [-1, 197, 768] | 0 |
| Linear | [-1, 197, 3072] | 2362368 | Attention | [-1, 197, 768] | 0 |
| GELU | [-1, 197, 3072] | 0 | LayerNorm | [-1, 197, 768] | 1536 |
| Dropout | [-1, 197, 3072] | 0 | Linear | [-1, 197, 3072] | 2362368 |
| Linear | [-1, 197, 768] | 2360064 | GELU | [-1, 197, 3072] | 0 |
| Dropout | [-1, 197, 768] | 0 | Dropout | [-1, 197, 3072] | 0 |
| MLP | [-1, 197, 768] | 0 | Linear | [-1, 197, 768] | 2360064 |
| Block | [-1, 197, 768] | 0 | Dropout | [-1, 197, 768] | 0 |
| LayerNorm | [-1, 197, 768] | 1536 | MLP | [-1, 197, 768] | 0 |
| Linear | [-1, 197, 2304] | 1771776 | Block | [-1, 197, 768] | 0 |
| Identity | [-1, 12, 197, 64] | 0 | LayerNorm | [-1, 768] | 1536 |
| Linear | [-1, 197, 768] | 590592 | Identity | [-1, 768] | 0 |
| Dropout | [-1, 197, 768] | 0 | Dropout | [-1, 3] | 0 |
| Attention | [-1, 197, 768] | 0 | Linear | [-1, 197, 768] | 2304 |

## 2.5. CNN

Convolutional neural networks (CNNs) are a powerful class of DL models [27] specifically designed for tasks involving image data [28]. In the case of histopathological image classification, CNNs offer a robust mechanism for automatically extracting and learning important features from the raw image data, such as textures, edges, and patterns [29]. These features are crucial for distinguishing between different types of tumor cells in medical images. Below, we explain the details of the CNN model used in this work, followed by its advantages and limitations. In this paper, we have implemented the CNN architecture to classify histopathological images into four distinct categories: NT, NVT, VT, and NVR.

The architecture starts with three blocks of convolutional layers. The first block contains two convolutional layers, each with 64 filters and a kernel size of (3, 3). LeakyReLU activation functions with an alpha value of 0.25 follow the convolutional layers, which allows a small gradient for negative inputs, preventing neurons from "dying" and improving the robustness of the model. This block ends with a MaxPooling2D layer that reduces the spatial size of the feature maps, focusing on the most important features while reducing computational complexity. The second block follows the same pattern, but with two convolutional layers using 128 filters, further enhancing the model's ability to capture more abstract features from the image data. The third convolutional block increases the number of filters to 256, helping the model learn even more detailed and complex features from the images. We used a learning rate of $10^{-4}$, determined through hyperparameter tuning, and the Adam optimizer for faster convergence and better adaptability during training. We then applied the GlobalAveragePooling2D layer, which reduces each feature map to a single value by taking the average across spatial dimensions. This technique reduces the number of parameters, making the fully connected layers more efficient and less prone to overfitting. The final layers consist of two fully connected (dense) layers with 1024 units each, both followed by LeakyReLU activations. The final layer in the network—a dense output layer with four units—represents the four output classes. It employs a SoftMax activation function to transform the raw output scores into probabilities, guaranteeing that the predicted class aligns with the highest probability. For training, we used a batch size of 32 and 30 epochs, which allowed the model to effectively classify histopathological images into four categories. We optimized the model's hyperparameters using the grid search method to balance extraction, computational efficiency, and generalization while addressing challenges like overfitting and computational costs.

## 2.6. CNN+ViT model (a hybrid model)

In this model, we have concatenated the features extracted from the CNN and Vision Transformer (ViT) into a single feature vector before passing them to the Multi-Layer Perceptron (MLP) as a well-known classifier [30] for classification. This process is crucial for integrating local features (captured by CNN) and global features (captured by ViT) into a unified representation, allowing the model to leverage both types of information for improved classification accuracy.

### 2.6.1. Extracting Features from CNN and ViT

CNN Features: After passing the input image through CNN, we obtain a feature vector of size 1024. The CNN uses a learning rate of $10^{-4}$, Adam optimizer, batch size of 32, and trains for 30 epochs, ensuring optimal local feature extraction.

ViT Features: Similarly, after processing the same input image through ViT, we extract a much larger feature vector of size 150,528. The ViT model uses Adam optimizer with parameters β1 = 0.9 and β2 = 0.999, a batch size of 32, and trains for 30 epochs, allowing it to capture global patterns effectively.

### 2.6.2. Concatenation of Features

Once we have both sets of features, we concatenate them into a single, unified feature vector of size 151,552. This concatenation combines the local and global features from CNN and ViT respectively.

The Multi-Layer Perceptron (MLP), with its fully connected layers, receives the concatenated 151,552-dimensional feature vector. The MLP processes this combined feature vector to classify the image into one of the four categories: NT, NVT, VT, and NVR. The MLP applies transformations to the vector using hidden layers and LeakyReLU activation functions with early stopping and evaluates the performance based on validation loss to prevent overfitting. Combining the detailed CNN features with the global ViT features can improve the model's classification accuracy by leveraging the strengths of both architectures.

Table 3. The architecture of our hybrid model for the classification of OS patients

| Stage/Layer | Layer Type | Input Shape | Output Shape | Description |
| --- | --- | --- | --- | --- |
| Input | Image | (128, 128, 3) | (128, 128, 3) | The input is an image of size 128x128 with 3 color channels (RGB). |
| CNN Feature Extraction | CNN (multiple conv & pooling) | (128, 128, 3) | (1024) | The CNN extracts 1024 local features from the image through several convolutional and pooling layers, capturing low-level and mid-level patterns. |
| ViT Feature Extraction | Vision Transformer (ViT) | (128, 128, 3) | (150,528) | The Vision Transformer divides the image into patches, applies self-attention, and extracts 150,528 global features, capturing contextual relationships. |
| Feature Concatenation | Concatenation | (1024) + (150,528) | (151,552) | The features extracted from CNN and ViT are concatenated to form a combined feature vector of size 151,552. |
| MLP - Layer 1 | Fully Connected Layer (Linear) | (151,552) | (1024) | The first fully connected layer reduces the feature vector to 1024 units, applying non-linearity (ReLU) for transformation. |
| MLP - Layer 2 | Fully Connected Layer (Linear) | (1024) | (256) | The second fully connected layer further reduces the dimensionality to 256 units, followed by ReLU activation. |
| Output Layer | Fully Connected Layer (Linear) | (256) | (4) | The final fully connected layer outputs 4 class logits (NT, NVT, VT, and NVR). |
| Softmax Layer | Softmax Activation | (4) | (4) | Convert the logits into probabilities for the four classes. |

All steps of the presented method including image processing, feature extraction methods, classification, and how to combine them for the intended four-class classification, are shown in Figure 3.

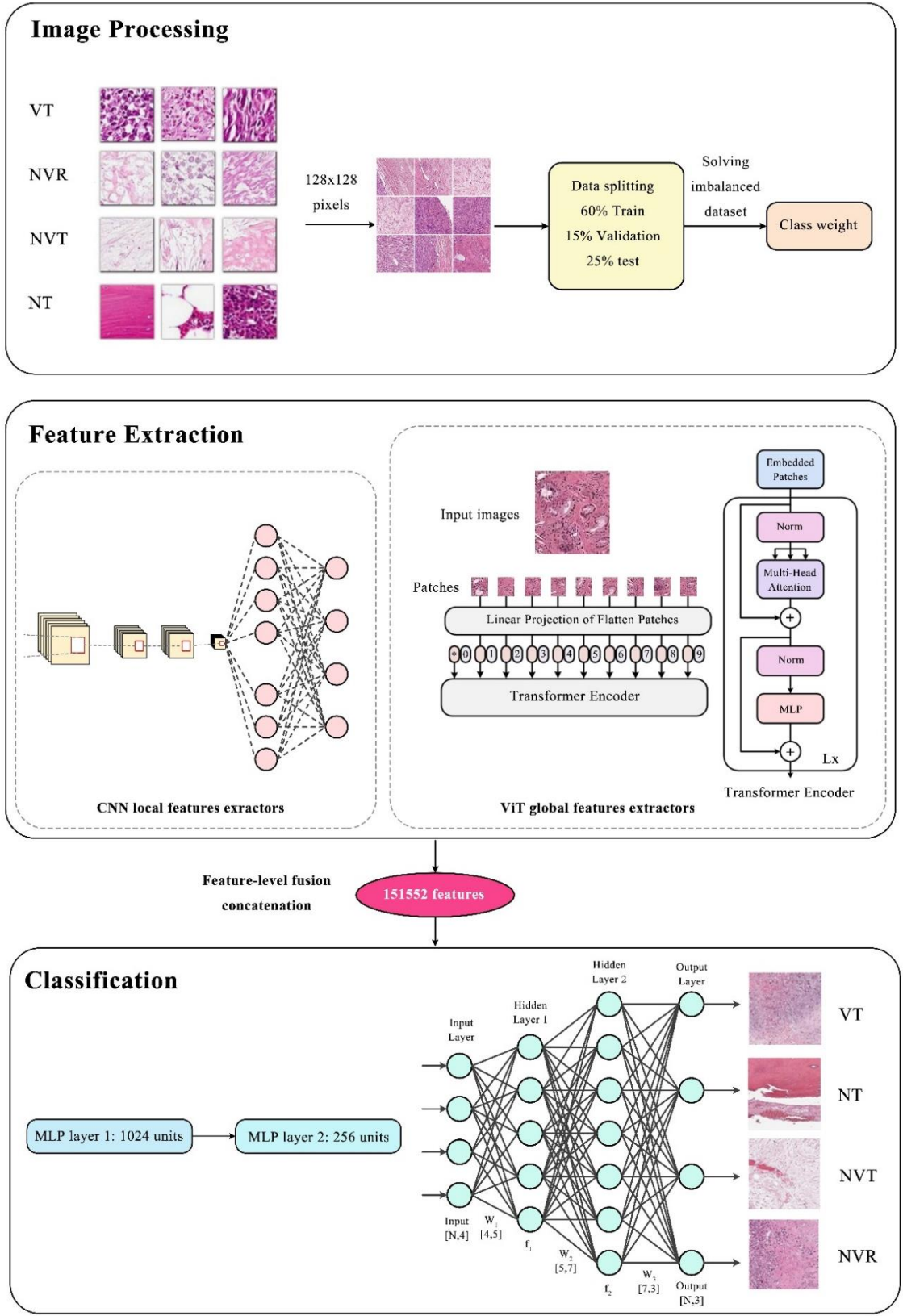

Figure 3: The overall procedure of the proposed hybrid method.

### 2.7. Evaluation metrics

The classification results are evaluated using accuracy, recall, precision, and F1-score:

$$Accuracy = \frac{TP+TN}{TP+FP+FN+TN} \quad (1)$$

$$Precision = \frac{TP}{TP+FP} \quad (2)$$

$$Recall = \frac{TP}{TP+FN} \quad (3)$$

$$F_1\_score = 2 \times \frac{precision \times recall}{precision + recall} \quad (4)$$

TP and TN are the numbers of true positives and negatives that are accurately labeled. FP and FN are incorrectly labeled samples [31].

### 3. Results and comparative study

In this study, we have evaluated the performance of four DL models—CNN, ViT, ResNet, and a hybrid model—for the classification of OS histopathological images. We have applied these models to datasets for two-class (the results are shown in Table 4), three-class (the results are shown in Table 5), and four-class classification tasks (the results are shown in Table 6). We have assessed each model's performance using key evaluation metrics such as accuracy, precision, recall, and F1-score (Table 3).

The CNN model showed decent performance, especially in two-class and three-class tasks. Specifically, for two-class classification, CNN achieved 82% test accuracy and 86% validation accuracy, demonstrating its capacity for binary classification tasks. It showed improvement in the three-class task, with 89% test accuracy and 90% validation accuracy. However, as the classification problem became more complex in the four-class task, CNN's performance dropped to 81% test accuracy and 88% validation accuracy.

On the other hand, ViT model uses its self-attention mechanism to capture global patterns, enabling it to model relationships between distant image regions. We observed that the ViT model consistently outperformed CNN in all tasks. For two-class classification, the ViT achieved 93% test accuracy and 94% validation accuracy, significantly better than the CNN. It also performed well in the three-class task, with 95% test accuracy and 93% validation accuracy. In the four-class task, ViT outperformed CNN again with 89% test accuracy and 88% validation accuracy (Table 2).

ResNet also performed well in these tasks. The two-class classification results showed 87% test accuracy and 92% validation accuracy, which is better than CNN but slightly below ViT. For the three-class task, ResNet achieved 87% test accuracy and 92% validation accuracy. However, in the four-class task, ResNet managed 87% test accuracy and 92% validation accuracy, reflecting a stable but less impressive performance compared to ViT (Table 3).

Among the tested models, the CNN + ViT hybrid model demonstrated the best performance across all tasks. In the two-class task, it achieved 93.67% test accuracy and 91.60% validation accuracy, outperforming both standalone CNN and ViT models. In the three-class task, the hybrid model was especially impressive, achieving a near-perfect 99.56% test accuracy and 99.91% validation accuracy, which sets a new benchmark for this task. For the four-class classification, which is the most challenging task, the hybrid model again outperformed all others, with 99.08% test accuracy and 99.70% validation accuracy.

Table 4. Results related to the binary Classification

| Model | Test Accuracy | Test Precision | Test Recall | Test F1-Score |
|---|---|---|---|---|
| CNN | 0.82 | 0.76 | 0.83 | 0.78 |
| ViT | 0.93 | 0.93 | 0.83 | 0.86 |
| CNN + ViT | **0.9367** | **0.9401** | **0.9367** | **0.9283** |
| ResNet | 0.87 | 0.65 | 0.69 | 0.67 |

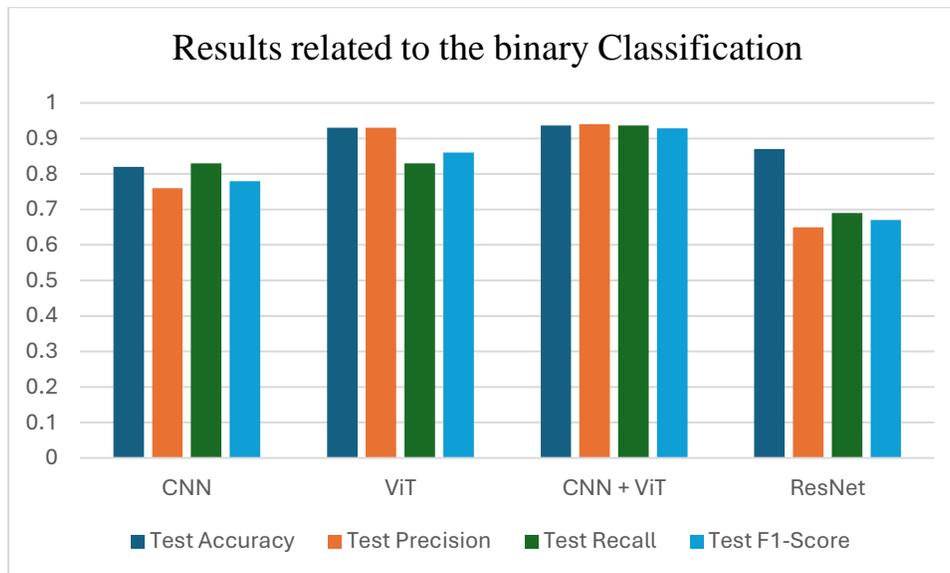

Figure 4: Results obtained by applying all classification models for VT vs. NT.

Figure 4 shows the average performance criteria for each of the models whose results are given in Table 4 for VT vs. NT patients.

Table 5. Results related to the three-Class Classification

| Model | Test Accuracy | Test Precision | Test Recall | Test F1-Score |
|---|---|---|---|---|
| CNN | 0.89 | 0.87 | 0.88 | 0.87 |
| ViT | 0.95 | 0.94 | 0.95 | 0.96 |
| CNN + ViT | **0.9956** | **0.9958** | **0.9990** | **0.9926** |
| ResNet | 0.87 | 0.82 | 0.87 | 0.84 |

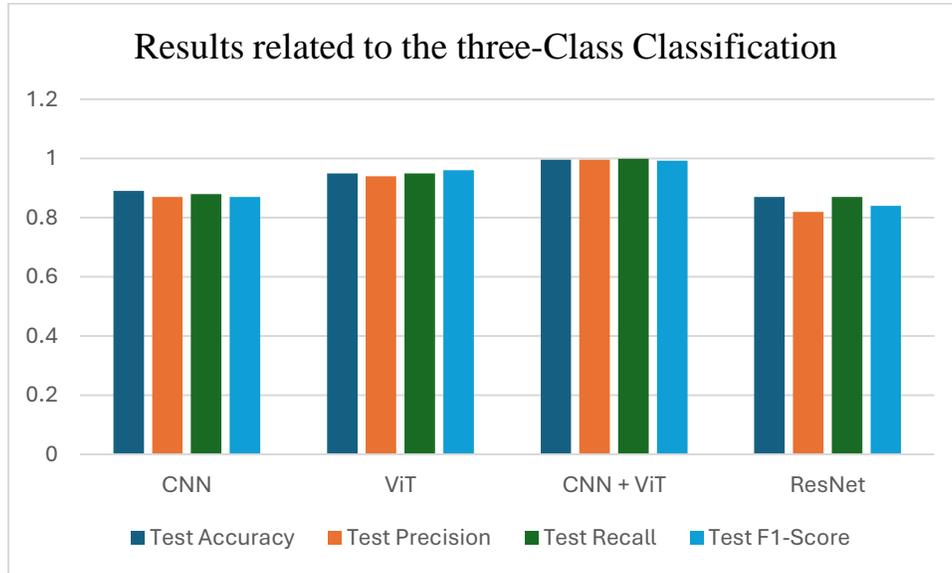

Figure 5: Results obtained by all classification models for VT vs. NVT vs. NT.

Figure 5 shows the average performance criteria for each of the models whose results are given in Table 5 for VT vs. NVT vs. NT patients.

Table 6: Results related to the four-Class Classification

| Model | Test Accuracy | Test Precision | Test Recall | Test F1-Score |
|---|---|---|---|---|
| CNN | 0.81 | 0.79 | 0.81 | 0.79 |
| ViT | 0.89 | 0.86 | 0.89 | 0.87 |
| CNN + ViT | **0.9908** | **0.9910** | **0.9928** | **0.9923** |
| ResNet | 0.87 | 0.82 | 0.87 | 0.84 |

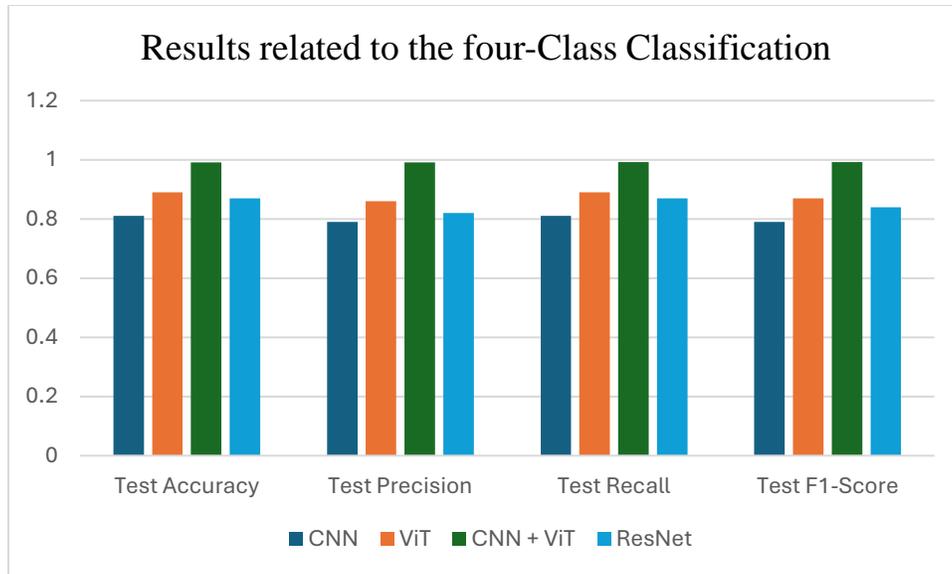

Figure 6: Results obtained by applying all classification models for VT vs. NVT vs. NT vs. NVR.

Figure 6 shows the average performance criteria for each of the models whose results are given in Table 6 for all patients.

### 3.1. Discussion

In this study, we developed a hybrid DL model that combines CNN and ViT to classify OS histopathological images into four distinct categories: NT, NVT, VT, and NVR. Our approach introduced several advantages over previous methods, improving classification accuracy and reducing computational complexity.

First, we achieved a higher accuracy (93.67%) in distinguishing VT from NT groups, surpassing previous studies that used the same dataset. For example, Mishra et al. [32] employed a custom CNN model, achieving an accuracy of 84% in binary classification between VT and NT. Similarly, another study [9] that used segmentation techniques reported a testing accuracy of 86% for the same task. Our model exceeded this benchmark, confirming the strength of the CNN + ViT hybrid architecture in effectively capturing both local and global image features with no segmentation step required.

Second, for the three-class classification (VT, NVT, NT), our model demonstrated an accuracy of 99.56% and a recall measure of 99.9% while reducing computational complexity. Prior research, such as that by Ahmed et al. [33], utilized two CNN models to classify three tumor types, achieving an accuracy of 86%. Fakieh et al. [34] improved upon this with a Wind Driven Optimization (WDO) and Deep Stacked Sparse Autoencoder (DSSAE) model, reaching an average accuracy of 99.71%. Moreover, Vaiyapuri et al. [35] developed a model incorporating the Honey Badger Optimization (HBO) algorithm and achieved an accuracy of 99.71%, with a high F1-score of 99.62%. Although these methods achieved high accuracy, they involved optimization techniques which increased computational demands. Our model, in contrast, maintained similar levels of accuracy but with reduced reliance on complex optimization methods, offering greater efficiency.

Furthermore, our work is the first to classify all four classes of OS (NT, NVT, VT, and NVR) using the TCGA dataset, setting a new benchmark for OS classification. We have achieved an accuracy of 99.08% in four-class classification.

Previous research primarily focused on binary or three-class classification, overlooking the critical NVR category. For instance, Mishra et al. [36] achieved an accuracy of 92.4% in three-class classification (VT, NVT, NT), but no prior studies had expanded to four-class classification. The non-viable ratio (NVR) category is crucial in osteosarcoma classification because it helps evaluate the effectiveness of treatment, particularly chemotherapy. A higher NVR, indicating a larger proportion of necrotic (non-viable) tissue compared to viable tumor areas, often correlates with a positive response to therapy and better patient outcomes. The NVR serves as a valuable prognostic tool, aiding clinicians in assessing tumor regression and adjusting treatment plans accordingly. Additionally, including the NVR in classifications offers a more comprehensive view of tumor heterogeneity.

Previous studies have often excluded the NVR category due to the complexity of accurately identifying and segmenting non-viable tissue in histopathological images. Traditional classification models focused primarily on simpler distinctions, such as viable versus non-viable tumor, which were more straightforward to define. Challenges like the lack of labeled data, segmentation difficulties, and the tendency to focus on clinically easier classifications limited the incorporation of NVR into earlier models. By addressing this gap in the research, our study's inclusion of the NVR category sets a new benchmark, offering more detailed insights into treatment outcomes and improving the accuracy of osteosarcoma classifications.

However, we observed that CNN's classification performance dropped in the four-class task. This suggests that while CNN excels at detecting local patterns, it struggles with tasks like specific image classification or segmentation activities that demand the model recognize and interpret larger, more complex relationships or structures within the image. This requires the model to understand broader, more complex image structures. We also observed that ResNet performance was lower than ViT for the four-class classification task. We think the reason is that ResNet can deal with problems like vanishing gradients due to its residual connections, but it still uses convolutional layers for feature extraction, which cannot model global information as well as ViT. The CNN + ViT hybrid model demonstrated the best performance among all models used for all classification tasks. These results clearly highlight the strength of combining the local feature extraction abilities of CNN with the global feature understanding capabilities of ViT.

Table 7. Related works on the classification of OS patients using TCIA dataset

| Related works | Classification methods | Classification type | The classes | Classification results |
|---|---|---|---|---|
| [32] 2017 | A Custom CNN | Binary | VT vs. NT | F1-score: 86% <br> Accuracy: 84% |
| [9] 2017 | A combination of image segmentation and analysis techniques | Binary | VT vs. NT | Testing accuracy of 86% |
| [37] 2018 | A Custom CNN | 3-Classes | VT vs. NVT vs. NT | Accuracy: 92.4% <br> Precision: 97% <br> Recall: 94% <br> F1-Score: 95% |
| [33] 2021 | Two CNN models | 3-Classes | VT vs. NVT vs. NT | Testing accuracy: 86% |
| [34] 2022 | The Wind Driven Optimization (WDO) and Deep Stacked Sparse Autoencoder (DSSAE) | 3-Classes | VT vs. NVT vs. NT | Average accuracy: 99.71% <br> Average precision: 99.28% <br> Average recall: 99.67% <br> Average F1-Score: 99.47% |
| [35] 2022 | Honey Badger Optimization with DL-Based Automated OS Classification (HBODL-AOC) model | 3-Classes | VT vs. NVT vs. NT | Accuracy: 99.71% <br> Precision: 99.57% <br> Recall: 99.68% <br> F1-Score: 99.62% <br> AUC Score: 99.73% |

In Table 7, we have shown the results of the articles for the classification OS patients using TCIA dataset.

### 4. Conclusion

The proposed hybrid AI model demonstrated significant improvements in accurately identifying critical features in OS images by merging the local feature extraction capabilities of CNNs with the global feature recognition strengths of ViTs.

This hybrid architecture outperformed traditional models, such as ResNet, by effectively leveraging both local and global features. The success of this model demonstrates the potential of combining these two approaches to advance medical imaging and improve personalized cancer treatment. Our results show that this hybrid method can significantly enhance diagnostic precision, streamline decision-making, and improve patient outcomes for diagnosis of OS patients.

Future research could focus on optimizing the model to reduce its computational demands, incorporating cancer-specific pre-training, and applying it to other cancer types. Testing this method in real-world clinical settings will be essential to assess its practicality and robustness for use in diagnosis and treatment planning.

**Acknowledgement**

**Limitations:** Not applicable.

**Funding for this study:**

This study was funded by ACMIT COMET Module FFG project (FFG number: 879733, application number: 39955962) In addition, this study is supported by FTI Dissertation grant (Project number: FTI23-D-037).

**Imaging Technique**

Histopathological image classification using deep learning

**Author Contributions**

Arezoo Borji was responsible for the programming, coding, method application, study design, data collection, analysis, and manuscript drafting. Dr. Hatamikia and Dr. Kronreif contributed to the conceptualization, supervision, interpretation of the analysis, and the review and editing of the manuscript. Dr. Angermayr contributed to the study's conceptualization and manuscript review.